\documentstyle[11pt,psfig,twoside]{article}

\begin{document}

\newcommand{\dd}{\,{\rm d}}
\newcommand{\ie}{{\it i.e.},\,}
\newcommand{\etal}{{\it et al.\ }}
\newcommand{\eg}{{\it e.g.},\,}
\newcommand{\cf}{{\it cf.\ }}
\newcommand{\vs}{{\it vs.\ }}
\newcommand{\zdot}{\makebox[0pt][l]{.}}
\newcommand{\up}[1]{\ifmmode^{\rm #1}\else$^{\rm #1}$\fi}
\newcommand{\dn}[1]{\ifmmode_{\rm #1}\else$_{\rm #1}$\fi}
\newcommand{\upd}{\up{d}}
\newcommand{\uph}{\up{h}}
\newcommand{\upm}{\up{m}}
\newcommand{\ups}{\up{s}}
\newcommand{\arcd}{\ifmmode^{\circ}\else$^{\circ}$\fi}
\newcommand{\arcm}{\ifmmode{'}\else$'$\fi}
\newcommand{\arcs}{\ifmmode{''}\else$''$\fi}
\newcommand{\MS}{{\rm M}\ifmmode_{\odot}\else$_{\odot}$\fi}
\newcommand{\RS}{{\rm R}\ifmmode_{\odot}\else$_{\odot}$\fi}
\newcommand{\LS}{{\rm L}\ifmmode_{\odot}\else$_{\odot}$\fi}

\newcommand{\Abstract}[2]{{\footnotesize\begin{center}ABSTRACT\end{center}
\vspace{1mm}\par#1\par
\noindent
{~}{\it #2}}}

\newcommand{\TabCap}[2]{\begin{center}\parbox[t]{#1}{\begin{center}
  \small {\spaceskip 2pt plus 1pt minus 1pt T a b l e}
  \refstepcounter{table}\thetable \\[2mm]
  \footnotesize #2 \end{center}}\end{center}}

\newcommand{\TableSep}[2]{\begin{table}[p]\vspace{#1}
\TabCap{#2}\end{table}}

\newcommand{\FigCap}[1]{\footnotesize\par\noindent Fig.\  %
  \refstepcounter{figure}\thefigure. #1\par}

\newcommand{\TableFont}{\footnotesize}
\newcommand{\TableFontIt}{\ttit}
\newcommand{\SetTableFont}[1]{\renewcommand{\TableFont}{#1}}

\newcommand{\MakeTable}[4]{\begin{table}[htb]\TabCap{#2}{#3}
  \begin{center} \TableFont \begin{tabular}{#1} #4 
  \end{tabular}\end{center}\end{table}}

\newcommand{\MakeTableSep}[4]{\begin{table}[p]\TabCap{#2}{#3}
  \begin{center} \TableFont \begin{tabular}{#1} #4 
  \end{tabular}\end{center}\end{table}}

\newenvironment{references}%
{
\footnotesize \frenchspacing
\renewcommand{\thesection}{}
\renewcommand{\in}{{\rm in }}
\renewcommand{\AA}{Astron.\ Astrophys.}
\newcommand{\AAS}{Astron.~Astrophys.~Suppl.~Ser.}
\newcommand{\ApJ}{Astrophys.\ J.}
\newcommand{\ApJS}{Astrophys.\ J.~Suppl.~Ser.}
\newcommand{\ApJL}{Astrophys.\ J.~Letters}
\newcommand{\AJ}{Astron.\ J.}
\newcommand{\IBVS}{IBVS}
\newcommand{\PASP}{P.A.S.P.}
\newcommand{\Acta}{Acta Astron.}
\newcommand{\MNRAS}{MNRAS}
\renewcommand{\and}{{\rm and }}
\section{{\rm REFERENCES}}
\sloppy \hyphenpenalty10000
\begin{list}{}{\leftmargin1cm\listparindent-1cm
\itemindent\listparindent\parsep0pt\itemsep0pt}}%
{\end{list}\vspace{2mm}}

\def\TYLDA{~}
\newlength{\DW}
\settowidth{\DW}{0}
\newcommand{\dw}{\hspace{\DW}}

\newcommand{\refitem}[5]{\item[]{#1} #2%
\def\REFARG{#3}\ifx\REFARG\TYLDA\else, {\it#3}\fi
\def\REFARG{#4}\ifx\REFARG\TYLDA\else, {\bf#4}\fi
\def\REFARG{#5}\ifx\REFARG\TYLDA\else, {#5}\fi.}

\newcommand{\Section}[1]{\section{#1}}
\newcommand{\Subsection}[1]{\subsection{#1}}
\newcommand{\Acknow}[1]{\par\vspace{5mm}{\bf Acknowledgements.} #1}
\pagestyle{myheadings}

\def\thefootnote{\fnsymbol{footnote}}
\begin{center}

{\Large\bf The Optical Gravitational Lensing Experiment.\\
\vskip3pt
Cepheids in Star Clusters from the Magellanic Clouds\footnote{Based 
on observations obtained with the 1.3~m Warsaw telescope at 
the Las Campanas Observatory operated by the Carnegie Institution of 
Washington.}}

\vskip 1cm

{ G.~~P~i~e~t~r~z~y~{\'n}~s~k~i,~~and~~A.~~U~d~a~l~s~k~i}
\vskip5mm
{Warsaw University Observatory, Al.~Ujazdowskie~4, 00-478~Warszawa,
Poland\\
e-mail: (pietrzyn,udalski)@astrouw.edu.pl\\}

\end{center}

\Abstract{
We present Cepheids located in the close neighborhood of star  clusters
from the Magellanic Clouds. 204 and 132 such stars were found in the  LMC
and SMC, respectively. The lists of objects were constructed based on 
catalogs of Cepheids and star clusters, recently published by the
OGLE-II  collaboration. Location of selected Cepheids on the sky  
indicates that many of them are very likely cluster  members.
Photometric data of Cepheids and clusters are available from the OGLE 
archive.}{~} 

\Section{Introduction}
Cepheids and star clusters are very important objects for empirical testing 
many fundamental problems in astronomy. Since the period-luminosity relation 
for Cepheids has been discovered by Leavitt (1912) they became one of the most 
important sources of information about distances in the nearby Universe. 
Observations of Cepheids also provide empirical constraints on theory of 
stellar structure, evolution etc. On the other hand star clusters are ideal 
tracers of stellar evolution and independent source of information about 
distances, ages, chemical composition, interstellar absorption etc.\ of the 
galaxies where they reside. 

Cepheids belonging to star clusters are especially worth detailed studies. For 
instance, due to well defined evolutionary phase they provide precise 
information on age, superior to that obtained with the standard procedure of 
isochrone fitting. They also may help in studies of cluster dynamics. 

Unfortunately the number of Cepheids located in the regions of star clusters 
is still small in both the Magellanic Clouds and the Galaxy. Observations are 
highly inhomogeneous, obtained by many astronomers, using many different 
instruments. Microlensing surveys make it possible to select large number of 
variable stars, in particular those from the fields of star clusters. 
Following the list of 127 eclipsing systems in optical coincidence with star 
clusters from the SMC (Pietrzy{\'n}ski and Udalski 1999), in this paper we 
present Cepheids located in the regions of star clusters in the Magellanic 
Clouds. 

\vspace*{-5pt}
\Section{Observational Data}
\vskip-5pt
The photometric data used in this paper were collected during the OGLE-II 
microlensing survey with the 1.3~m Warsaw telescope located at the Las 
Campanas Observatory, Chile, which is operated by the Carnegie Institution of 
Washington. The telescope was equipped with ${2048\times2048}$ CCD camera 
working in driftscan mode. Detailed description of the instrumental system and 
the OGLE-II project was presented by Udalski, Kubiak and Szyma{\'n}ski (1997). 

About 4.5 and 2.4 square degrees regions in the LMC and SMC, respectively, 
covering most of the bars of these galaxies were monitored regularly since 
January 1997 through the standard {\it BVI} filters. Coordinates of the 
observed fields and the schematic maps of the LMC and SMC with contours of 
observed fields can be found in Udalski \etal (1999c,d). Data reduction 
pipeline and data quality tests of the SMC photometry are described in Udalski 
\etal (1998). Quality of the LMC data is similar and it will be described with 
release of stellar maps of the LMC in the near future (Udalski \etal in 
preparation). In particular, accuracy of transformation to the standard system 
is about 0.01--0.02~mag for all {\it BVI}-bands. 

\vspace*{-5pt}
\Section{Cepheids in the Magellanic Cloud Star Clusters}
\vspace*{-5pt}
Tables~1 and 2 list Cepheids located in the close neighborhood of star 
clusters of the LMC and SMC, respectively. The lists were constructed based on 
Catalogs of Star Clusters from the LMC (Pietrzy{\'n}ski \etal 1999) and SMC 
(Pietrzy{\'n}ski \etal 1998) and Catalogs of Cepheids from the LMC and SMC 
(Udalski \etal 1999c,d). 
\renewcommand{\TableFont}{\tiny}
\MakeTableSep{l@{\hspace{5pt}}
           l@{\hspace{3pt}}
           r@{\hspace{3pt}}
           c@{\hspace{3pt}}
           r@{\hspace{3pt}}
           c@{\hspace{3pt}}
           c@{\hspace{3pt}}
           c@{\hspace{3pt}}
           c@{\hspace{3pt}}
           c@{\hspace{3pt}}
}{13cm}
{Cepheids in star clusters from the LMC}
{
\hline
\noalign{\vskip2pt}
\multicolumn{1}{c}{Cluster name} & \multicolumn{1}{c}{Field} & 
\multicolumn{1}{c}{ID} & D & \multicolumn{1}{c}{$P$}& $T_0{-}2450000$ & 
$V$ & $I$ & $E(B-V)$ & Type \\
\noalign{\vskip1pt}
\multicolumn{1}{c}{OGLE-CL-} & & & $[R_{\rm CL}]$ & 
\multicolumn{1}{c}{[days]} & [HJD] & [mag] & [mag] & [mag] & \\
\noalign{\vskip2pt}
\hline
\noalign{\vskip3pt}
LMC0005 & LMC\_SC15 &  45780 &   0.7 &    3.58788 &  723.23710 &  16.017 &  15.227 &    0.126 &  FU \\ 
LMC0038 & LMC\_SC14 & 109715 &   0.1 &    1.91300 &  723.27982 &  15.898 &  15.337 &    0.138 &  FO \\ 
LMC0051 & LMC\_SC14 & 160642 &   0.3 &    0.97211 &  724.94922 &  15.348 &  15.140 &    0.138 &  BR \\ 
LMC0052 & LMC\_SC14 & 143866 &   0.4 &    2.04587 &  724.55632 &  16.136 &  15.391 &    0.142 &  FO \\ 
LMC0054 & LMC\_SC14 & 170005 &   0.5 &   20.64642 &  712.44747 &  14.347 &  13.344 &    0.138 &  FA \\ 
LMC0058 & LMC\_SC14 & 220934 &   0.6 &    0.86912 &  724.83583 &  16.982 &  16.409 &    0.142 &  DM \\ 
LMC0058 & LMC\_SC14 & 221178 &   1.2 &   16.74144 &  717.42796 &  18.870 &  17.666 &    0.142 &  FA \\ 
LMC0069 & LMC\_SC12 & 200768 &   1.2 &   14.87020 &  716.87361 &  16.716 &  15.790 &    0.124 &  FA \\ 
LMC0074 & LMC\_SC12 &  37723 &   0.6 &    2.47581 &  444.05113 &  17.318 &  16.894 &    0.139 &  FA \\ 
LMC0090 & LMC\_SC13 & 111968 &   1.1 &    8.33337 &  722.56247 &  14.975 &  14.138 &    0.135 &  FU \\ 
LMC0093 & LMC\_SC13 & 111968 &   0.4 &    8.33337 &  722.56247 &  14.975 &  14.138 &    0.135 &  FU \\ 
LMC0113 & LMC\_SC13 & 173734 &   0.7 &   12.72593 &  713.03927 &  14.609 &  13.694 &    0.135 &  FU \\ 
LMC0115 & LMC\_SC13 & 178831 &   0.8 &   17.45740 &  713.31458 &  14.256 &  13.276 &    0.135 &  FU \\ 
LMC0125 & LMC\_SC11 & 228660 &   0.8 &    4.20424 &  722.18621 &  15.774 &  14.952 &    0.129 &  FU \\ 
LMC0136 & LMC\_SC11 & 118714 &   0.8 &    3.09884 &  723.48360 &  16.251 &  15.486 &    0.154 &  FU \\ 
LMC0142 & LMC\_SC11 & 250872 &   0.2 &   18.65508 &  717.25663 &  13.964 &  13.056 &    0.152 &  FU \\ 
LMC0142 & LMC\_SC11 & 250925 &   0.0 &    5.56688 &  719.67472 &  14.477 &  13.879 &    0.152 &  FO \\ 
LMC0142 & LMC\_SC11 & 250938 &   0.2 &    8.55890 &  719.65413 &  14.650 &  13.913 &    0.153 &  FU \\ 
LMC0152 & LMC\_SC11 & 257240 &   1.4 &   11.86131 &  713.17954 &  14.422 &  13.607 &    0.152 &  FU \\ 
LMC0154 & LMC\_SC11 & 338308 &   0.9 &    1.74277 &  723.93868 &  16.266 &  15.609 &    0.152 &  FO \\ 
LMC0156 & LMC\_SC11 & 338308 &   1.3 &    1.74277 &  723.93868 &  16.266 &  15.609 &    0.152 &  FO \\ 
LMC0157 & LMC\_SC11 & 306838 &   0.8 &    1.06752 &  724.41740 &  17.333 &  16.534 &    0.151 &  FO \\ 
LMC0164 & LMC\_SC11 &  49743 &   1.3 &    2.29575 &  443.10180 &  15.753 &  15.080 &    0.146 &  FO \\ 
LMC0164 & LMC\_SC11 &  49799 &   0.8 &    1.56520 &  444.92204 &  16.307 &  15.635 &    0.146 &  FO \\ 
LMC0164 & LMC\_SC11 &  54641 &   0.5 &    2.71388 &  444.56352 &  16.192 &  15.452 &    0.132 &  FU \\ 
LMC0164 & LMC\_SC11 & 331546 &   1.3 &   11.90536 &  713.18698 &  14.626 &  13.720 &    0.152 &  FU \\ 
LMC0164 & LMC\_SC11 & 331601 &   0.7 &   56.49819 &  686.12410 &  15.116 &  14.959 &    0.152 &  FA \\ 
LMC0209 & LMC\_SC10 & 250332 &   0.6 &    2.01253 &  443.28253 &  15.858 &  15.185 &    0.147 &  FO \\ 
LMC0211 & LMC\_SC10 & 245266 &   0.8 &    3.96572 &  443.23028 &  15.559 &  14.852 &    0.147 &  FU \\ 
LMC0261 & LMC\_SC8 &  76174 &   0.4 &    2.90619 &  442.43915 &  15.308 &  14.658 &    0.142 &  FO \\ 
LMC0261 & LMC\_SC8 &  76176 &   0.9 &    2.46017 &  444.78291 &  15.480 &  14.812 &    0.142 &  FO \\ 
LMC0261 & LMC\_SC8 &  76179 &   0.3 &    2.46242 &  443.09468 &  15.611 &  14.925 &    0.142 &  FO \\ 
LMC0262 & LMC\_SC8 &  86096 &   0.4 &    3.82230 &  444.32857 &  15.900 &  15.081 &    0.142 &  FU \\ 
LMC0263 & LMC\_SC8 &  21319 &   1.2 &    5.42859 &  440.73741 &  15.146 &  14.440 &    0.133 &  FU \\ 
LMC0266 & LMC\_SC8 &  39745 &   0.9 &    6.89000 &  441.24680 &  15.308 &  14.463 &    0.133 &  FU \\ 
LMC0269 & LMC\_SC8 &  64734 &   0.3 &    3.53975 &  443.36233 &  16.039 &  15.263 &    0.136 &  FU \\ 
LMC0276 & LMC\_SC8 & 145110 &   0.5 &    3.44680 &  443.93145 &  15.978 &  15.267 &    0.136 &  FU \\ 
LMC0278 & LMC\_SC8 & 118595 &   0.3 &    6.45160 &  440.43021 &  15.224 &  14.430 &    0.133 &  FU \\ 
LMC0285 & LMC\_SC8 & 151151 &   0.6 &    2.03362 &  444.28845 &  15.945 &  15.322 &    0.136 &  FO \\ 
LMC0303 & LMC\_SC8 & 242825 &   0.1 &    1.15426 &  444.15257 &  16.736 &  16.130 &    0.136 &  DM \\ 
LMC0304 & LMC\_SC8 & 363511 &   0.3 &    0.92234 &  444.30269 &  17.218 &  16.550 &    0.142 &  FO \\ 
LMC0305 & LMC\_SC8 & 298699 &   1.0 &    4.79323 &  444.01190 &  15.410 &  14.642 &    0.133 &  FU \\ 
LMC0306 & LMC\_SC8 & 337664 &   0.2 &    0.62980 &  444.55248 &  17.165 &  16.672 &    0.136 &  DM \\ 
LMC0309 & LMC\_SC8 & 318671 &   0.9 &    8.17005 &  443.08395 &  14.579 &  13.852 &    0.133 &  FU \\ 
LMC0312 & LMC\_SC8 & 318671 &   0.9 &    8.17005 &  443.08395 &  14.579 &  13.852 &    0.133 &  FU \\ 
LMC0318 & LMC\_SC7 &  21841 &   1.0 &    4.81308 &  442.24754 &  15.580 &  14.816 &    0.143 &  FU \\ 
LMC0321 & LMC\_SC7 &  30199 &   0.4 &    4.10536 &  441.92826 &  15.709 &  14.977 &    0.138 &  FU \\ 
LMC0321 & LMC\_SC7 &  30200 &   0.1 &    3.28632 &  442.74251 &  15.755 &  15.121 &    0.138 &  FU \\ 
LMC0332 & LMC\_SC7 & 192212 &   0.2 &    2.41249 &  444.85037 &  15.620 &  15.036 &    0.142 &  FO \\ 
LMC0336 & LMC\_SC7 & 174573 &   0.9 &    3.76557 &  444.39097 &  16.445 &  15.540 &    0.138 &  FU \\ 
LMC0344 & LMC\_SC7 & 325360 &   0.6 &    2.64961 &  443.25720 &  16.343 &  15.643 &    0.142 &  FU \\ 
LMC0350 & LMC\_SC7 & 311535 &   1.4 &    4.32379 &  444.14237 &  15.111 &  14.373 &    0.142 &  FO \\ 
LMC0362 & LMC\_SC7 &  86027 &   0.5 &    0.40354 &  444.87352 &  18.111 &  17.563 &    0.107 &  FO \\ 
LMC0362 & LMC\_SC7 & 425296 &   0.5 &    0.40354 &  444.87720 &  18.112 &  17.537 &    0.142 &  FO \\ 
LMC0365 & LMC\_SC21&  40943 &   1.4 &    1.80312 &  724.92207 &  16.054 &  15.385 &    0.146 &  FO \\ 
LMC0375 & LMC\_SC6 &  66641 &   0.2 &    2.71388 &  444.00445 &  16.111 &  15.483 &    0.107 &  FU \\ 
LMC0379 & LMC\_SC6 &  66530 &   1.4 &    8.69372 &  442.84074 &  14.706 &  13.966 &    0.107 &  FU \\ 
LMC0394 & LMC\_SC6 & 254054 &   0.3 &    3.09206 &  442.57315 &  16.391 &  15.577 &    0.138 &  FU \\ 
LMC0394 & LMC\_SC6 & 254057 &   0.4 &    3.38313 &  443.41709 &  16.519 &  15.613 &    0.138 &  FU \\ 
LMC0394 & LMC\_SC6 & 254530 &   1.5 &    1.26602 &  444.93725 &  18.249 &  17.751 &    0.138 &  FA \\ 
LMC0395 & LMC\_SC6 & 254054 &   0.9 &    3.09206 &  442.57315 &  16.391 &  15.577 &    0.138 &  FU \\ 
LMC0395 & LMC\_SC6 & 254057 &   1.3 &    3.38313 &  443.41709 &  16.519 &  15.613 &    0.138 &  FU \\ 
LMC0395 & LMC\_SC6 & 254530 &   1.2 &    1.26602 &  444.93725 &  18.249 &  17.751 &    0.138 &  FA \\ 
LMC0401 & LMC\_SC6 & 267410 &   0.3 &    0.88863 &  444.49611 &  17.232 &  16.588 &    0.126 &  FO \\ 
LMC0407 & LMC\_SC6 & 422348 &   0.7 &    5.63875 &  443.91906 &  15.306 &  14.581 &    0.107 &  FU \\ 
LMC0408 & LMC\_SC6 & 431558 &   0.4 &    2.24761 &  444.31615 &  15.579 &  15.007 &    0.107 &  FO \\ 
LMC0410 & LMC\_SC6 &  26913 &   0.2 &    5.73854 &  444.31113 &  15.559 &  14.622 &    0.115 &  FU \\ 
LMC0410 & LMC\_SC6 & 377026 &   0.2 &    5.73913 &  444.22041 &  15.513 &  14.627 &    0.138 &  FU \\
}

\setcounter{table}{0}
\MakeTableSep{l@{\hspace{5pt}}
           l@{\hspace{3pt}}
           r@{\hspace{3pt}}
           c@{\hspace{3pt}}
           r@{\hspace{3pt}}
           c@{\hspace{3pt}}
           c@{\hspace{3pt}}
           c@{\hspace{3pt}}
           c@{\hspace{3pt}}
           c@{\hspace{3pt}}
}{13cm}
{continued}
{
\hline
\noalign{\vskip2pt}
\multicolumn{1}{c}{Cluster name} & \multicolumn{1}{c}{Field} & 
\multicolumn{1}{c}{ID} & D & \multicolumn{1}{c}{$P$}& $T_0{-}2450000$ & 
$V$ & $I$ & $E(B-V)$ & Type \\
\noalign{\vskip1pt}
\multicolumn{1}{c}{OGLE-CL-} & & & $[R_{\rm CL}]$ & 
\multicolumn{1}{c}{[days]} & [HJD] & [mag] & [mag] & [mag] & \\
\noalign{\vskip2pt}
\hline
\noalign{\vskip3pt}
LMC0411 & LMC\_SC21 &     12 &   0.1 &    3.23227 &  442.70858 &  16.226 &  15.323 &    0.130 &  FU \\ 
LMC0411 & LMC\_SC21 & 187786 &   0.2 &    3.85993 &  723.67241 &  15.929 &  15.066 &    0.146 &  FU \\ 
LMC0411 & LMC\_SC21 & 187788 &   0.2 &    2.81419 &  722.79915 &  15.636 &  15.081 &    0.146 &  FU \\ 
LMC0411 & LMC\_SC21 & 187792 &   0.1 &    4.52988 &  724.83766 &  15.797 &  14.927 &    0.146 &  FU \\ 
LMC0411 & LMC\_SC21 & 187797 &   0.6 &    2.97310 &  723.45071 &  16.460 &  15.488 &    0.146 &  FU \\ 
LMC0411 & LMC\_SC21 & 187840 &   0.6 &    2.01501 &  724.55164 &  16.050 &  15.339 &    0.146 &  FO \\ 
LMC0411 & LMC\_SC21 & 187849 &   0.1 &    3.28019 &  724.58073 &  16.187 &  15.261 &    0.146 &  FU \\ 
LMC0411 & LMC\_SC21 & 187853 &   0.1 &    3.35487 &  724.73951 &  16.604 &  15.598 &    0.146 &  FU \\ 
LMC0411 & LMC\_SC21 & 187856 &   0.3 &    3.16725 &  723.96832 &  16.067 &  15.201 &    0.146 &  FU \\ 
LMC0412 & LMC\_SC6 & 422324 &   0.7 &    2.23825 &  443.72420 &  15.785 &  15.135 &    0.107 &  FO \\ 
LMC0431 & LMC\_SC5 & 275320 &   0.1 &    3.66130 &  444.10632 &  14.994 &  14.356 &    0.115 &  FO \\ 
LMC0431 & LMC\_SC5 & 275412 &   1.5 &    2.37496 &  443.48611 &  16.361 &  15.668 &    0.115 &  FU \\ 
LMC0436 & LMC\_SC5 & 372083 &   1.0 &    5.67416 &  442.21699 &  15.474 &  14.629 &    0.115 &  FU \\ 
LMC0438 & LMC\_SC5 & 399066 &   0.7 &    5.40730 &  442.52121 &  15.393 &  14.579 &    0.115 &  FU \\ 
LMC0457 & LMC\_SC4 &  53463 &   0.5 &    5.39550 &  440.52924 &  15.308 &  14.392 &    0.120 &  FO \\ 
LMC0457 & LMC\_SC4 &  53514 &   1.3 &    5.97551 &  442.02593 &  15.533 &  14.642 &    0.120 &  FU \\ 
LMC0457 & LMC\_SC4 &  53518 &   0.1 &    5.56684 &  443.02577 &  15.930 &  14.998 &    0.120 &  FU \\ 
LMC0457 & LMC\_SC4 &  53527 &   0.2 &    7.68466 &  440.55810 &  16.029 &  14.834 &    0.120 &  FU \\ 
LMC0457 & LMC\_SC4 &  53528 &   0.3 &    9.37703 &  440.78678 &  16.417 &  15.003 &    0.120 &  FU \\ 
LMC0457 & LMC\_SC4 &  53546 &   1.2 &    6.40602 &  444.56575 &  15.237 &  14.435 &    0.120 &  FU \\ 
LMC0457 & LMC\_SC4 &  53796 &   0.5 &    1.29505 &  443.98043 &  17.610 &  16.614 &    0.120 &  DM \\ 
LMC0457 & LMC\_SC4 & 176266 &   1.4 &    8.80196 &  439.87795 &  15.249 &  14.295 &    0.120 &  FU \\ 
LMC0459 & LMC\_SC4 &  36266 &   1.2 &    6.80614 &  438.48105 &  15.512 &  14.563 &    0.120 &  FU \\ 
LMC0461 & LMC\_SC4 &  53463 &   1.4 &    5.39550 &  440.52924 &  15.308 &  14.392 &    0.120 &  FO \\ 
LMC0461 & LMC\_SC4 & 176266 &   1.5 &    8.80196 &  439.87795 &  15.249 &  14.295 &    0.120 &  FU \\ 
LMC0461 & LMC\_SC4 & 176307 &   1.2 &    2.97632 &  444.48446 &  15.265 &  14.637 &    0.120 &  FO \\ 
LMC0466 & LMC\_SC4 & 167947 &   0.3 &    4.68002 &  443.46227 &  15.743 &  14.949 &    0.120 &  FU \\ 
LMC0468 & LMC\_SC4 & 220148 &   1.3 &    0.74088 &  444.47941 &  17.327 &  16.731 &    0.118 &  DM \\ 
LMC0477 & LMC\_SC4 & 296000 &   0.3 &    4.13355 &  443.50669 &  15.629 &  14.939 &    0.120 &  FU \\ 
LMC0478 & LMC\_SC4 & 296023 &   0.2 &    3.54545 &  444.94859 &  15.904 &  15.194 &    0.120 &  FU \\ 
LMC0480 & LMC\_SC4 & 295932 &   1.1 &    4.16628 &  441.63659 &  14.996 &  14.338 &    0.120 &  FO \\ 
LMC0482 & LMC\_SC4 & 305691 &   0.8 &    7.45743 &  443.64407 &  14.970 &  14.173 &    0.105 &  FU \\ 
LMC0487 & LMC\_SC4 & 417848 &   1.3 &    5.88305 &  442.73212 &  15.114 &  14.373 &    0.120 &  FU \\ 
LMC0490 & LMC\_SC4 & 391255 &   0.1 &    3.47904 &  442.50816 &  15.435 &  14.667 &    0.120 &  FO \\ 
LMC0491 & LMC\_SC4 & 408742 &   0.3 &    7.07075 &  438.18394 &  14.996 &  14.208 &    0.120 &  FU \\ 
LMC0495 & LMC\_SC4 & 408738 &   0.5 &    7.32026 &  443.12864 &  14.831 &  14.028 &    0.120 &  FU \\ 
LMC0496 & LMC\_SC3 &  35233 &   0.4 &    4.01553 &  444.36757 &  14.970 &  14.265 &    0.120 &  FO \\ 
LMC0507 & LMC\_SC3 & 170205 &   1.4 &    7.57042 &  444.30099 &  15.135 &  14.284 &    0.120 &  FU \\ 
LMC0508 & LMC\_SC3 & 201460 &   1.1 &    2.72805 &  443.37541 &  15.621 &  14.943 &    0.123 &  FO \\ 
LMC0512 & LMC\_SC3 & 170246 &   0.0 &    2.62452 &  444.94181 &  15.567 &  14.913 &    0.120 &  FO \\ 
LMC0512 & LMC\_SC3 & 170248 &   0.2 &    5.29412 &  443.84860 &  15.546 &  14.717 &    0.120 &  FU \\ 
LMC0514 & LMC\_SC3 & 250776 &   0.8 &    4.61320 &  440.42176 &  16.794 &  15.582 &    0.134 &  FU \\ 
LMC0527 & LMC\_SC3 & 368172 &   0.6 &    4.92028 &  444.23791 &  15.639 &  14.807 &    0.120 &  FU \\ 
LMC0532 & LMC\_SC2 &  47348 &   1.3 &    6.29131 &  444.72622 &  15.066 &  14.313 &    0.121 &  FU \\ 
LMC0533 & LMC\_SC3 &  47348 &   0.9 &    6.29131 &  444.72622 &  15.066 &  14.313 &    0.121 &  FU \\ 
LMC0538 & LMC\_SC2 &  78835 &   0.9 &    2.58776 &  443.97512 &  16.129 &  15.456 &    0.151 &  FU \\ 
LMC0540 & LMC\_SC2 &  70874 &   1.3 &    2.75064 &  442.37314 &  15.496 &  14.842 &    0.150 &  FO \\ 
LMC0541 & LMC\_SC2 &  91891 &   1.2 &    4.16669 &  443.54047 &  15.770 &  14.991 &    0.131 &  FU \\ 
LMC0543 & LMC\_SC2 & 180132 &   0.1 &    7.09089 &  440.20200 &  15.172 &  14.329 &    0.150 &  FU \\ 
LMC0551 & LMC\_SC2 & 158664 &   0.5 &    7.02416 &  440.35802 &  15.184 &  14.383 &    0.121 &  FU \\ 
LMC0551 & LMC\_SC2 & 158669 &   1.1 &    3.04825 &  444.97874 &  15.523 &  14.823 &    0.121 &  FO \\ 
LMC0552 & LMC\_SC2 & 127941 &   0.3 &    2.36507 &  443.41876 &  15.557 &  14.984 &    0.121 &  FO \\ 
LMC0556 & LMC\_SC2 & 232835 &   0.6 &    4.63835 &  441.82571 &  15.676 &  14.901 &    0.121 &  FU \\ 
LMC0559 & LMC\_SC2 & 263427 &   1.4 &    3.30708 &  444.08152 &  15.973 &  15.273 &    0.121 &  FU \\ 
LMC0559 & LMC\_SC2 & 263433 &   0.3 &    3.23673 &  442.04624 &  15.375 &  14.710 &    0.121 &  FO \\ 
LMC0559 & LMC\_SC2 & 263498 &   0.7 &    3.88073 &  443.61059 &  16.104 &  15.269 &    0.121 &  FU \\ 
LMC0565 & LMC\_SC2 & 240715 &   1.1 &    2.52449 &  444.07778 &  16.366 &  15.672 &    0.121 &  FU \\ 
LMC0565 & LMC\_SC2 & 240852 &   1.5 &    0.80429 &  444.25721 &  17.318 &  16.722 &    0.121 &  FO \\ 
LMC0565 & LMC\_SC2 & 334165 &   1.1 &    3.29488 &  442.81323 &  16.171 &  15.350 &    0.121 &  FU \\ 
LMC0566 & LMC\_SC2 & 334165 &   1.3 &    3.29488 &  442.81323 &  16.171 &  15.350 &    0.121 &  FU \\ 
LMC0569 & LMC\_SC2 & 357821 &   0.5 &    4.29598 &  441.30282 &  15.797 &  14.973 &    0.121 &  FU \\ 
LMC0571 & LMC\_SC1 &  51886 &   1.4 &    4.65494 &  442.59661 &  15.638 &  14.826 &    0.147 &  FU \\ 
LMC0574 & LMC\_SC1 &  66584 &   0.9 &    1.83677 &  444.77986 &  16.443 &  15.678 &    0.147 &  FO \\ 
LMC0585 & LMC\_SC1 & 150925 &   0.1 &    4.29163 &  443.64129 &  15.893 &  15.055 &    0.147 &  FU \\ 
LMC0585 & LMC\_SC1 & 150939 &   1.4 &    2.36133 &  444.75922 &  15.790 &  15.124 &    0.147 &  FO \\ 
LMC0591 & LMC\_SC1 & 158021 &   0.2 &    2.93704 &  444.49429 &  16.079 &  15.372 &    0.147 &  DM \\ 
LMC0591 & LMC\_SC1 & 158027 &   1.4 &    2.28500 &  444.56718 &  16.020 &  15.284 &    0.147 &  FO \\ 
LMC0591 & LMC\_SC1 & 158066 &   1.4 &    2.20349 &  443.00294 &  16.102 &  15.365 &    0.147 &  FO \\ 
}
\setcounter{table}{0}
\MakeTableSep{l@{\hspace{5pt}}
           l@{\hspace{3pt}}
           r@{\hspace{3pt}}
           c@{\hspace{3pt}}
           r@{\hspace{3pt}}
           c@{\hspace{3pt}}
           c@{\hspace{3pt}}
           c@{\hspace{3pt}}
           c@{\hspace{3pt}}
           c@{\hspace{3pt}}
}{13cm}
{concluded}
{
\hline
\noalign{\vskip2pt}
\multicolumn{1}{c}{Cluster name} & \multicolumn{1}{c}{Field} & 
\multicolumn{1}{c}{ID} & D & \multicolumn{1}{c}{$P$}& $T_0{-}2450000$ & 
$V$ & $I$ & $E(B-V)$ & Type \\
\noalign{\vskip1pt}
\multicolumn{1}{c}{OGLE-CL-} & & & $[R_{\rm CL}]$ & 
\multicolumn{1}{c}{[days]} & [HJD] & [mag] & [mag] & [mag] & \\
\noalign{\vskip2pt}
\hline
\noalign{\vskip3pt}
LMC0592 & LMC\_SC1 & 164361 &   1.0 &    1.88935 &  444.79581 &  16.414 &  15.641 &    0.163 &  FO \\ 
LMC0599 & LMC\_SC1 & 201683 &   0.9 &    2.31922 &  444.82428 &  15.870 &  15.174 &    0.152 &  FO \\ 
LMC0603 & LMC\_SC1 & 306814 &   1.3 &    2.23221 &  444.53573 &  15.883 &  15.192 &    0.147 &  FO \\ 
LMC0603 & LMC\_SC1 & 306872 &   1.1 &    3.55525 &  442.84736 &  16.065 &  15.252 &    0.147 &  FU \\ 
LMC0607 & LMC\_SC1 & 266530 &   0.0 &    4.11529 &  441.55075 &  15.541 &  14.857 &    0.117 &  FU \\ 
LMC0609 & LMC\_SC1 & 324972 &   0.5 &    4.65729 &  443.11470 &  15.467 &  14.732 &    0.147 &  FU \\ 
LMC0620 & LMC\_SC16 &  37107 &   0.3 &    2.85225 &  722.24340 &  15.524 &  14.859 &    0.185 &  FO \\ 
LMC0622 & LMC\_SC16 &  26114 &   0.7 &    2.45436 &  723.55163 &  15.646 &  15.015 &    0.148 &  FO \\ 
LMC0622 & LMC\_SC16 &  99253 &   0.7 &    3.21026 &  723.51146 &  15.854 &  15.197 &    0.148 &  FU \\ 
LMC0622 & LMC\_SC16 &  99255 &   0.2 &    3.63447 &  724.98569 &  15.888 &  15.073 &    0.148 &  FU \\ 
LMC0622 & LMC\_SC16 &  99257 &   0.1 &    2.25531 &  724.55848 &  15.665 &  15.083 &    0.148 &  FO \\ 
LMC0622 & LMC\_SC16 &  99259 &   1.2 &    2.54551 &  724.46521 &  15.552 &  14.921 &    0.148 &  FO \\ 
LMC0622 & LMC\_SC16 &  99294 &   0.3 &    2.05606 &  723.61242 &  15.770 &  15.181 &    0.148 &  FO \\ 
LMC0626 & LMC\_SC16 & 104483 &   0.7 &    4.18017 &  724.95320 &  15.809 &  15.013 &    0.148 &  FU \\ 
LMC0627 & LMC\_SC16 & 115249 &   0.6 &    4.32412 &  720.90521 &  16.211 &  15.224 &    0.185 &  FU \\ 
LMC0627 & LMC\_SC16 & 115254 &   1.4 &    3.88261 &  724.25392 &  16.064 &  15.247 &    0.185 &  FU \\ 
LMC0631 & LMC\_SC16 & 172438 &   1.5 &    3.67029 &  724.88064 &  16.237 &  15.423 &    0.148 &  FU \\ 
LMC0631 & LMC\_SC16 & 172452 &   0.2 &    2.12037 &  724.14454 &  15.798 &  15.212 &    0.148 &  FO \\ 
LMC0631 & LMC\_SC16 & 177829 &   1.5 &    2.70457 &  723.74610 &  16.295 &  15.554 &    0.185 &  FU \\ 
LMC0632 & LMC\_SC16 & 177811 &   0.1 &    4.88909 &  722.06980 &  15.515 &  14.761 &    0.185 &  FU \\ 
LMC0633 & LMC\_SC16 & 172383 &   0.1 &    5.33924 &  722.54152 &  15.592 &  14.776 &    0.148 &  FU \\ 
LMC0633 & LMC\_SC16 & 172435 &   0.5 &    2.27102 &  723.10424 &  15.964 &  15.285 &    0.148 &  FO \\ 
LMC0633 & LMC\_SC16 & 172447 &   0.4 &    3.64388 &  723.94776 &  15.832 &  15.154 &    0.148 &  FU \\ 
LMC0633 & LMC\_SC16 & 172450 &   0.2 &    2.12318 &  724.80143 &  15.955 &  15.307 &    0.148 &  FO \\ 
LMC0633 & LMC\_SC16 & 172455 &   0.1 &    1.97267 &  723.60901 &  15.894 &  15.295 &    0.148 &  FO \\ 
LMC0633 & LMC\_SC16 & 172459 &   0.2 &    2.08792 &  724.43182 &  15.889 &  15.276 &    0.148 &  FO \\ 
LMC0633 & LMC\_SC16 & 172460 &   0.1 &    3.34067 &  723.46521 &  15.989 &  15.249 &    0.148 &  FU \\ 
LMC0633 & LMC\_SC16 & 177773 &   0.7 &    4.14658 &  723.88892 &  15.933 &  15.054 &    0.185 &  FU \\ 
LMC0633 & LMC\_SC16 & 177774 &   0.2 &    4.97643 &  723.99702 &  15.496 &  14.761 &    0.185 &  FU \\ 
LMC0633 & LMC\_SC16 & 177777 &   0.3 &    4.67694 &  723.82548 &  15.476 &  14.792 &    0.185 &  FU \\ 
LMC0633 & LMC\_SC16 & 177781 &   0.8 &    5.34543 &  724.04817 &  15.580 &  14.800 &    0.185 &  FU \\ 
LMC0633 & LMC\_SC16 & 177823 &   0.3 &    1.91792 &  724.12452 &  16.033 &  15.415 &    0.185 &  FO \\ 
LMC0633 & LMC\_SC16 & 235480 &   0.8 &    2.04557 &  724.33725 &  16.041 &  15.367 &    0.148 &  FO \\ 
LMC0633 & LMC\_SC16 & 240725 &   1.3 &   17.56780 &  716.34020 &   9.999 &  17.212 &    0.185 &  FA \\ 
LMC0634 & LMC\_SC16 & 167363 &   0.8 &    4.72921 &  723.64560 &  15.599 &  14.849 &    0.148 &  FU \\ 
LMC0635 & LMC\_SC16 & 194262 &   0.3 &    2.14536 &  723.91789 &  16.551 &  15.619 &    0.181 &  FO \\ 
LMC0636 & LMC\_SC16 & 240459 &   1.0 &    5.57865 &  724.29562 &  15.574 &  14.702 &    0.185 &  FU \\ 
LMC0636 & LMC\_SC16 & 240469 &   0.5 &    3.30523 &  721.83339 &  15.636 &  14.862 &    0.185 &  FO \\ 
LMC0636 & LMC\_SC16 & 240517 &   1.1 &    2.23872 &  724.40841 &  15.982 &  15.290 &    0.185 &  FO \\ 
LMC0636 & LMC\_SC16 & 240518 &   0.8 &    2.12913 &  724.77135 &  16.041 &  15.347 &    0.185 &  FO \\ 
LMC0636 & LMC\_SC16 & 240524 &   0.2 &    2.21601 &  723.11858 &  15.831 &  15.212 &    0.185 &  FO \\ 
LMC0636 & LMC\_SC16 & 240525 &   0.5 &    1.86071 &  723.50445 &  16.006 &  15.419 &    0.185 &  FO \\ 
LMC0638 & LMC\_SC16 & 257336 &   1.1 &    2.40550 &  723.83795 &  16.613 &  15.552 &    0.181 &  FO \\ 
LMC0648 & LMC\_SC17 &  33286 &   0.4 &    2.21517 &  723.84675 &  15.829 &  15.157 &    0.175 &  FO \\ 
LMC0648 & LMC\_SC17 &  33289 &   0.4 &    4.82058 &  721.57077 &  15.522 &  14.760 &    0.175 &  FU \\ 
LMC0648 & LMC\_SC17 &  33290 &   0.2 &    2.55444 &  722.78060 &  15.864 &  15.110 &    0.175 &  DM \\ 
LMC0648 & LMC\_SC17 &  33292 &   0.3 &    4.39425 &  722.62297 &  15.833 &  14.977 &    0.175 &  FU \\ 
LMC0648 & LMC\_SC17 &  33296 &   0.4 &    2.33253 &  723.97765 &  15.842 &  15.135 &    0.176 &  FO \\ 
LMC0648 & LMC\_SC17 &  33299 &   0.6 &    4.04385 &  722.43556 &  15.921 &  15.035 &    0.175 &  FU \\ 
LMC0648 & LMC\_SC17 &  33301 &   0.9 &    5.78371 &  722.82435 &  15.839 &  14.842 &    0.175 &  FU \\ 
LMC0648 & LMC\_SC17 &  33306 &   0.9 &    3.36728 &  723.64425 &  16.318 &  15.423 &    0.176 &  FU \\ 
LMC0648 & LMC\_SC17 &  33351 &   0.8 &    1.85403 &  724.42809 &  16.059 &  15.347 &    0.175 &  FO \\ 
LMC0648 & LMC\_SC17 &  33368 &   0.2 &    3.64424 &  724.56633 &  15.956 &  15.145 &    0.175 &  FU \\ 
LMC0650 & LMC\_SC17 &  45207 &   0.4 &    3.17309 &  723.74520 &  16.245 &  15.390 &    0.175 &  FU \\ 
LMC0656 & LMC\_SC17 & 102941 &   0.3 &    4.48780 &  724.93418 &  15.886 &  15.067 &    0.175 &  FU \\ 
LMC0659 & LMC\_SC17 & 117748 &   0.8 &    3.57231 &  723.51809 &  15.652 &  14.815 &    0.201 &  FO \\ 
LMC0665 & LMC\_SC17 & 161761 &   0.2 &    2.94886 &  724.58682 &  16.444 &  15.620 &    0.175 &  FU \\ 
LMC0678 & LMC\_SC18 &  69137 &   0.2 &    4.18875 &  722.84863 &  16.345 &  15.346 &    0.178 &  FU \\ 
LMC0681 & LMC\_SC18 &  81454 &   1.0 &    2.02331 &  724.11582 &  16.256 &  15.484 &    0.173 &  FO \\ 
LMC0683 & LMC\_SC18 &  89202 &   0.9 &    1.47140 &  724.51654 &  17.011 &  16.244 &    0.173 &  DM \\ 
LMC0685 & LMC\_SC18 & 144662 &   0.6 &    2.38412 &  724.29685 &  15.956 &  15.221 &    0.173 &  FO \\ 
LMC0690 & LMC\_SC18 & 110903 &   0.1 &    3.08563 &  724.29846 &  16.534 &  15.613 &    0.182 &  FU \\ 
LMC0694 & LMC\_SC19 &  28796 &   0.5 &    2.39931 &  723.84475 &  16.546 &  15.740 &    0.187 &  FU \\ 
LMC0702 & LMC\_SC19 &  81235 &   0.4 &    3.60974 &  724.58983 &  16.239 &  15.357 &    0.187 &  FU \\ 
LMC0703 & LMC\_SC19 &  77859 &   0.9 &    3.61971 &  722.02986 &  16.087 &  15.262 &    0.187 &  FU \\ 
LMC0715 & LMC\_SC19 & 148467 &   1.2 &    2.85623 &  722.32707 &  16.422 &  15.583 &    0.153 &  FU \\ 
LMC0715 & LMC\_SC19 & 148475 &   0.2 &    2.97442 &  724.16899 &  16.233 &  15.454 &    0.153 &  FU \\ 
LMC0739 & LMC\_SC20 & 145017 &   1.3 &    2.66457 &  724.82090 &  16.352 &  15.541 &    0.142 &  FU \\ 
}

\renewcommand{\arraystretch}{1.02}
\setcounter{table}{1}
\MakeTableSep{l@{\hspace{5pt}}
           l@{\hspace{3pt}}
           r@{\hspace{3pt}}
           c@{\hspace{3pt}}
           r@{\hspace{3pt}}
           c@{\hspace{3pt}}
           c@{\hspace{3pt}}
           c@{\hspace{3pt}}
           c@{\hspace{3pt}}
           c@{\hspace{3pt}}
}{13cm}
{Cepheids in star clusters from the SMC}
{
\hline
\noalign{\vskip2pt}
\multicolumn{1}{c}{Cluster name} & \multicolumn{1}{c}{Field} & 
\multicolumn{1}{c}{ID} & D & \multicolumn{1}{c}{$P$}& $T_0{-}2450000$ & 
$V$ & $I$ & $E(B-V)$ & Type \\
\noalign{\vskip1pt}
\multicolumn{1}{c}{OGLE-CL-} & & & $[R_{\rm CL}]$ & 
\multicolumn{1}{c}{[days]} & [HJD] & [mag] & [mag] & [mag] & \\
\noalign{\vskip2pt}
\hline
\noalign{\vskip3pt}
SMC0007 & SMC\_SC2 &  41567 &    0.4 &    1.40886 &  618.74916 &    16.936 &    16.346 & 0.078 & FO\\ 
SMC0008 & SMC\_SC2 &  35276 &    0.2 &   17.46250 &  610.07361 &    14.720 &    13.819 & 0.078 & FU\\ 
SMC0009 & SMC\_SC2 &  35278 &    0.2 &   10.28820 &  618.88967 &    15.299 &    14.417 & 0.078 & FU\\ 
SMC0015 & SMC\_SC3 &  19968 &    1.4 &    6.22889 &  615.49328 &    15.964 &    15.247 & 0.089 & FU\\ 
SMC0016 & SMC\_SC3 &  28018 &    0.6 &    1.85081 &  619.56624 &    16.509 &    15.747 & 0.089 & FU\\ 
SMC0016 & SMC\_SC3 &  28026 &    0.8 &    3.24153 &  619.57762 &    16.502 &    15.879 & 0.089 & FU\\ 
SMC0016 & SMC\_SC3 &  28071 &    0.1 &    1.73099 &  618.87379 &    17.348 &    16.609 & 0.089 & FU\\ 
SMC0016 & SMC\_SC3 &  28166 &    0.5 &    1.61426 &  619.28998 &    17.618 &    16.925 & 0.089 & FU\\ 
SMC0016 & SMC\_SC3 &  28170 &    0.5 &    0.96090 &  619.64731 &    17.460 &    16.887 & 0.089 & FO\\ 
SMC0021 & SMC\_SC3 & 100628 &    0.9 &    1.69913 &  618.31681 &    17.210 &    16.593 & 0.089 & FU\\ 
SMC0024 & SMC\_SC3 & 226052 &    0.8 &    2.35922 &  619.87879 &    16.414 &    15.729 & 0.089 & FO\\ 
SMC0024 & SMC\_SC3 & 226093 &    0.9 &    1.67881 &  619.30658 &    17.008 &    16.437 & 0.089 & FU\\ 
SMC0024 & SMC\_SC3 & 226219 &    0.7 &    1.42663 &  618.92733 &    17.631 &    16.947 & 0.089 & FU\\ 
SMC0025 & SMC\_SC3 & 202814 &    1.3 &    1.35409 &  619.36886 &    17.147 &    16.497 & 0.089 & FO\\ 
SMC0026 & SMC\_SC3 & 193306 &    0.7 &    3.55879 &  616.75811 &    15.431 &    14.786 & 0.089 & FO\\ 
SMC0026 & SMC\_SC3 & 193346 &    0.9 &    1.35113 &  618.92071 &    17.353 &    16.619 & 0.089 & FO\\ 
SMC0027 & SMC\_SC3 & 184934 &    1.3 &    2.29836 &  618.48704 &    16.974 &    16.279 & 0.089 & FU\\ 
SMC0030 & SMC\_SC4 &  26050 &    1.2 &    2.24353 &  618.07520 &    17.082 &    16.256 & 0.094 & FU\\ 
SMC0032 & SMC\_SC4 &   2200 &    0.2 &    8.03958 &  618.41522 &    15.251 &    14.525 & 0.094 & FU\\ 
SMC0033 & SMC\_SC4 &  56758 &    0.4 &    8.49275 &  615.16825 &    15.734 &    14.805 & 0.094 & FU\\ 
SMC0034 & SMC\_SC4 &  75328 &    0.2 &    1.84920 &  619.40231 &    17.170 &    16.396 & 0.094 & FO\\ 
SMC0038 & SMC\_SC4 & 113676 &    0.4 &    1.84525 &  618.31789 &    17.293 &    16.631 & 0.094 & FU\\ 
SMC0039 & SMC\_SC4 & 101175 &    1.5 &    5.09495 &  617.96620 &    16.162 &    15.361 & 0.094 & FU\\ 
SMC0039 & SMC\_SC4 & 101254 &    0.3 &    2.12686 &  619.33069 &    17.085 &    16.412 & 0.094 & FU\\ 
SMC0042 & SMC\_SC4 & 103754 &    1.4 &    1.42543 &  618.67533 &    16.608 &    16.111 & 0.094 & FO\\ 
SMC0043 & SMC\_SC4 & 167207 &    1.2 &   11.77230 &  615.62092 &    15.444 &    14.393 & 0.094 & FU\\ 
SMC0043 & SMC\_SC4 & 167294 &    0.6 &    1.12408 &  619.63158 &    16.518 &    15.982 & 0.094 & SO\\ 
SMC0044 & SMC\_SC4 & 182573 &    1.3 &    3.36192 &  618.08487 &    15.819 &    15.135 & 0.094 & FO\\ 
SMC0044 & SMC\_SC4 & 182698 &    0.4 &    1.27799 &  619.48514 &    17.019 &    16.418 & 0.094 & FO\\ 
SMC0045 & SMC\_SC4 & 149863 &    0.0 &    2.10220 &  618.11212 &    16.724 &    15.966 & 0.094 & FU\\ 
SMC0045 & SMC\_SC4 & 149945 &    0.6 &    1.93618 &  618.33651 &    17.078 &    16.364 & 0.094 & FU\\ 
SMC0045 & SMC\_SC4 & 149961 &    0.2 &    1.70359 &  619.88353 &    17.159 &    16.571 & 0.094 & FU\\ 
SMC0046 & SMC\_SC4 & 149830 &    1.0 &    3.12054 &  618.25082 &    15.747 &    15.113 & 0.094 & FO\\ 
SMC0048 & SMC\_SC4 & 163672 &    0.5 &    0.57435 &  619.97776 &    18.330 &    17.705 & 0.094 & DM\\ 
SMC0048 & SMC\_SC5 &  21147 &    1.0 &    1.71604 &  464.05698 &    17.325 &    16.636 & 0.101 & FU\\ 
SMC0048 & SMC\_SC5 &  21488 &    0.5 &    0.57434 &  464.90685 &    18.305 &    17.691 & 0.101 & FO\\ 
SMC0048 & SMC\_SC5 & 160151 &    1.2 &    0.95653 &  619.48722 &    18.182 &    17.459 & 0.101 & FO\\ 
SMC0054 & SMC\_SC5 &  95232 &    0.9 &    1.62340 &  463.52647 &    16.436 &    15.870 & 0.101 & FO\\ 
SMC0054 & SMC\_SC5 &  95332 &    1.2 &    1.43148 &  464.63410 &    17.564 &    16.944 & 0.101 & FU\\ 
SMC0057 & SMC\_SC5 & 123315 &    0.4 &    9.93769 &  456.85612 &    15.375 &    14.495 & 0.101 & FU\\ 
SMC0057 & SMC\_SC5 & 123380 &    1.0 &    1.88441 &  464.34760 &    16.694 &    16.018 & 0.101 & FO\\ 
SMC0058 & SMC\_SC5 & 140700 &    1.3 &    2.75631 &  464.78389 &    16.774 &    16.011 & 0.101 & FU\\ 
SMC0058 & SMC\_SC5 & 140909 &    1.0 &    1.39567 &  464.36678 &    18.036 &    17.222 & 0.101 & FU\\ 
SMC0060 & SMC\_SC5 & 170190 &    1.3 &    3.38513 &  462.26488 &    15.732 &    15.107 & 0.101 & FO\\ 
SMC0063 & SMC\_SC5 & 202240 &    0.4 &    1.94635 &  464.87498 &    17.038 &    16.417 & 0.101 & FU\\ 
SMC0064 & SMC\_SC5 & 213952 &    0.6 &   27.41300 &  447.26716 &    14.176 &    13.178 & 0.101 & FU\\ 
SMC0064 & SMC\_SC5 & 213983 &    1.2 &    3.61449 &  461.66443 &     9.999 &    15.327 & 0.101 & FO\\ 
SMC0066 & SMC\_SC5 & 266150 &    1.0 &    1.58000 &  464.93150 &    17.499 &    16.823 & 0.101 & FU\\ 
SMC0066 & SMC\_SC5 & 271051 &    0.1 &   15.64740 &  451.81515 &    14.607 &    13.704 & 0.101 & FU\\ 
SMC0067 & SMC\_SC5 & 316843 &    0.7 &    1.11613 &  464.67251 &    17.220 &    16.630 & 0.101 & FO\\ 
SMC0068 & SMC\_SC5 & 260821 &    0.8 &    6.81805 &  461.30849 &    15.488 &    14.714 & 0.101 & FU\\ 
SMC0068 & SMC\_SC5 & 260976 &    1.4 &    1.55747 &  463.96344 &    17.170 &    16.603 & 0.101 & FU\\ 
SMC0068 & SMC\_SC5 & 260992 &    1.4 &    1.28816 &  463.83946 &    17.299 &    16.628 & 0.101 & FO\\ 
SMC0069 & SMC\_SC5 & 271164 &    0.3 &    2.30034 &  463.15391 &    16.886 &    16.157 & 0.101 & FU\\ 
SMC0071 & SMC\_SC6 & 288734 &    1.1 &    4.88882 &  463.95655 &    15.961 &    15.188 & 0.094 & FU\\ 
SMC0071 & SMC\_SC6 & 288872 &    1.4 &    1.46495 &  463.80656 &    17.775 &    17.042 & 0.094 & FU\\ 
SMC0074 & SMC\_SC6 &  49142 &    0.3 &    3.92114 &  463.79105 &    15.962 &    15.360 & 0.094 & FU\\ 
SMC0074 & SMC\_SC6 &  49153 &    0.6 &    2.72070 &  464.39482 &    16.325 &    15.531 & 0.094 & FO\\ 
SMC0074 & SMC\_SC6 &  49155 &    1.1 &    5.88439 &  464.86236 &    16.098 &    15.234 & 0.094 & FO\\ 
SMC0074 & SMC\_SC6 &  49197 &    0.9 &    2.58151 &  464.21633 &    16.479 &    15.874 & 0.094 & FU\\ 
SMC0074 & SMC\_SC6 &  49238 &    0.9 &    2.07911 &  463.23369 &    17.246 &    16.486 & 0.094 & FU\\ 
SMC0074 & SMC\_SC6 &  49351 &    0.1 &    2.07796 &  463.10747 &    16.192 &    15.690 & 0.094 & FO\\ 
SMC0075 & SMC\_SC6 &  29034 &    0.6 &    3.93931 &  463.57879 &    16.260 &    15.545 & 0.094 & FU\\ 
SMC0078 & SMC\_SC6 & 128740 &    0.2 &    9.88728 &  462.12553 &    15.358 &    14.495 & 0.094 & FU\\ 
SMC0078 & SMC\_SC6 & 128892 &    1.4 &    0.76210 &  464.27047 &    16.945 &    16.698 & 0.094 & FO\\ 
SMC0079 & SMC\_SC6 &  94519 &    0.4 &    3.11878 &  464.05801 &    16.743 &    16.003 & 0.094 & FU\\ 
}

\setcounter{table}{1}
\MakeTableSep{l@{\hspace{5pt}}
           l@{\hspace{3pt}}
           r@{\hspace{3pt}}
           c@{\hspace{3pt}}
           r@{\hspace{3pt}}
           c@{\hspace{3pt}}
           c@{\hspace{3pt}}
           c@{\hspace{3pt}}
           c@{\hspace{3pt}}
           c@{\hspace{3pt}}
}{13cm}
{concluded}
{
\hline
\noalign{\vskip2pt}
\multicolumn{1}{c}{Cluster name} & \multicolumn{1}{c}{Field} & 
\multicolumn{1}{c}{ID} & D & \multicolumn{1}{c}{$P$}& $T_0{-}2450000$ & 
$V$ & $I$ & $E(B-V)$ & Type \\
\noalign{\vskip1pt}
\multicolumn{1}{c}{OGLE-CL-} & & & $[R_{\rm CL}]$ & 
\multicolumn{1}{c}{[days]} & [HJD] & [mag] & [mag] & [mag] & \\
\noalign{\vskip2pt}
\hline
\noalign{\vskip3pt}
SMC0085 & SMC\_SC6 & 153030 &    1.2 &    2.13571 &  463.76086 &    17.331 &    16.596 & 0.094 & FU\\ 
SMC0087 & SMC\_SC6 &  89698 &    1.4 &    1.91914 &  463.17510 &    17.277 &    16.607 & 0.094 & FU\\ 
SMC0088 & SMC\_SC6 & 141593 &    0.5 &    5.72171 &  459.60445 &    15.721 &    14.933 & 0.094 & FU\\ 
SMC0088 & SMC\_SC6 & 141749 &    1.2 &    1.49391 &  464.23696 &    17.638 &    16.988 & 0.094 & FU\\ 
SMC0090 & SMC\_SC6 & 175773 &    1.2 &    2.31818 &  464.18010 &    16.253 &    15.516 & 0.094 & FO\\ 
SMC0090 & SMC\_SC6 & 180129 &    0.9 &    1.59418 &  464.19936 &    17.503 &    16.836 & 0.094 & FU\\ 
SMC0092 & SMC\_SC6 & 232227 &    0.6 &    2.67844 &  462.60605 &    16.201 &    15.493 & 0.094 & FO\\ 
SMC0093 & SMC\_SC6 & 242093 &    1.2 &    2.00387 &  464.11086 &    16.908 &    16.301 & 0.094 & FU\\ 
SMC0099 & SMC\_SC7 &  70819 &    0.9 &    2.87953 &  619.52783 &    15.974 &    15.359 & 0.097 & FU\\ 
SMC0103 & SMC\_SC7 &  91545 &    0.2 &    1.78021 &  619.30486 &    16.725 &    16.101 & 0.097 & FO\\ 
SMC0105 & SMC\_SC7 & 110088 &    0.5 &    6.09280 &  619.69447 &    15.868 &    15.037 & 0.097 & FU\\ 
SMC0105 & SMC\_SC7 & 110197 &    0.6 &    0.95703 &  619.44017 &    18.047 &    17.196 & 0.097 & FO\\ 
SMC0105 & SMC\_SC7 & 110251 &    1.5 &    1.70804 &  618.96889 &    17.660 &    16.937 & 0.097 & FU\\ 
SMC0107 & SMC\_SC7 & 206038 &    1.1 &    1.18480 &  619.77435 &    17.051 &    16.498 & 0.097 & FO\\ 
SMC0117 & SMC\_SC8 & 139531 &    1.0 &    1.67514 &  619.00859 &     9.999 &    16.983 & 0.100 & FU\\ 
SMC0118 & SMC\_SC8 & 204460 &    0.2 &    2.49360 &  617.62931 &    15.781 &    15.186 & 0.100 & FO\\ 
SMC0119 & SMC\_SC8 & 201506 &    0.7 &    5.37653 &  619.80231 &    15.893 &    15.093 & 0.100 & FU\\ 
SMC0120 & SMC\_SC8 & 201506 &    1.0 &    5.37653 &  619.80231 &    15.893 &    15.093 & 0.100 & FU\\ 
SMC0122 & SMC\_SC8 & 163502 &    0.6 &    3.48000 &  617.07068 &    16.570 &    15.763 & 0.100 & FO\\ 
SMC0122 & SMC\_SC8 & 163504 &    0.4 &    1.44514 &  619.24474 &    16.367 &    15.804 & 0.100 & FO\\ 
SMC0124 & SMC\_SC9 &  33090 &    0.9 &    0.67490 &  619.80537 &    17.929 &    17.393 & 0.076 & FO\\ 
SMC0127 & SMC\_SC9 &  86924 &    0.1 &    1.90782 &  618.16653 &    15.933 &    15.479 & 0.076 & FO\\ 
SMC0129 & SMC\_SC9 &  70439 &    1.2 &    0.78463 &  619.71184 &    17.288 &    16.841 & 0.076 & FO\\ 
SMC0130 & SMC\_SC9 &  89376 &    0.3 &    2.11771 &  619.34262 &    16.089 &    15.536 & 0.076 & FO\\ 
SMC0130 & SMC\_SC9 & 129534 &    1.1 &   18.11310 &  611.88043 &    14.440 &    13.519 & 0.076 & FU\\ 
SMC0132 & SMC\_SC9 &  94309 &    0.5 &    1.04715 &  619.14816 &    17.345 &    16.809 & 0.076 & FO\\ 
SMC0135 & SMC\_SC9 & 147031 &    0.2 &    0.66243 &  620.00063 &    17.877 &    17.307 & 0.076 & FO\\ 
SMC0137 & SMC\_SC10 &   8949 &    0.4 &    1.01211 &  619.92625 &    16.677 &    16.181 & 0.079 & FO\\ 
SMC0141 & SMC\_SC10 &  52210 &    0.9 &    1.76375 &  619.32186 &    16.289 &    15.725 & 0.079 & FO\\ 
SMC0141 & SMC\_SC10 &  52222 &    1.5 &    1.46101 &  618.65742 &    16.915 &    16.358 & 0.079 & FO\\ 
SMC0141 & SMC\_SC10 &  52269 &    0.2 &    2.35633 &  618.20938 &    17.166 &    16.437 & 0.079 & FU\\ 
SMC0149 & SMC\_SC10 & 108084 &    1.0 &    2.07929 &  618.05528 &    15.991 &    15.458 & 0.079 & FO\\ 
SMC0149 & SMC\_SC10 & 108101 &    0.6 &    3.12498 &  617.76822 &    16.303 &    15.612 & 0.079 & FU\\ 
SMC0149 & SMC\_SC10 & 108122 &    0.6 &    3.22932 &  619.75437 &    16.395 &    15.678 & 0.079 & FU\\ 
SMC0149 & SMC\_SC10 & 108123 &    0.8 &    2.40357 &  619.61427 &    16.552 &    15.925 & 0.079 & FU\\ 
SMC0151 & SMC\_SC10 & 114449 &    0.2 &    2.29254 &  619.27995 &    16.241 &    15.592 & 0.079 & FO\\ 
SMC0154 & SMC\_SC11 &  46557 &    1.3 &    2.93693 &  618.62552 &    16.127 &    15.535 & 0.084 & FU\\ 
SMC0154 & SMC\_SC11 &  46577 &    0.6 &    1.70625 &  619.98375 &    17.179 &    16.333 & 0.084 & FO\\ 
SMC0155 & SMC\_SC11 &  50956 &    0.4 &    5.91124 &  614.99409 &    15.419 &    14.784 & 0.084 & FU\\ 
SMC0156 & SMC\_SC11 &  40376 &    1.0 &    1.65571 &  618.64254 &    17.218 &    16.627 & 0.084 & FU\\ 
SMC0156 & SMC\_SC11 &  42383 &    1.2 &    0.80714 &  619.79471 &    17.703 &    17.128 & 0.084 & DM\\ 
SMC0157 & SMC\_SC11 &  32136 &    1.1 &    1.93392 &  618.19445 &    16.673 &    16.133 & 0.084 & FU\\ 
SMC0157 & SMC\_SC11 &  32190 &    0.3 &    1.17536 &  619.68682 &    17.745 &    17.148 & 0.084 & FU\\ 
SMC0158 & SMC\_SC11 &  87074 &    1.3 &    2.87814 &  618.26466 &    15.673 &    15.089 & 0.084 & FO\\ 
SMC0158 & SMC\_SC11 &  87075 &    0.7 &    3.20045 &  619.65570 &    16.305 &    15.607 & 0.084 & FU\\ 
SMC0158 & SMC\_SC11 &  89112 &    0.6 &    3.07455 &  618.83560 &    16.074 &    15.438 & 0.084 & FU\\ 
SMC0159 & SMC\_SC11 &  68073 &    0.7 &    0.91530 &  619.48105 &    17.613 &    17.014 & 0.084 & FO\\ 
SMC0159 & SMC\_SC11 &  70009 &    0.8 &    1.42400 &  619.81444 &    16.250 &    15.488 & 0.084 & BR\\ 
SMC0159 & SMC\_SC11 &  99428 &    1.1 &    1.83184 &  618.92241 &    16.935 &    16.258 & 0.084 & FU\\ 
SMC0171 & SMC\_SC2 &  78238 &    0.4 &    0.64958 &  619.91618 &    18.100 &    17.531 & 0.078 & FO\\ 
SMC0179 & SMC\_SC3 & 217715 &    1.2 &    1.45733 &  619.54705 &    17.636 &    16.947 & 0.089 & FU\\ 
SMC0189 & SMC\_SC4 & 167230 &    0.4 &    3.90049 &  617.53356 &    16.788 &    15.889 & 0.094 & FU\\ 
SMC0190 & SMC\_SC4 & 192867 &    0.6 &    3.69281 &  616.72477 &    16.061 &    15.418 & 0.094 & FU\\ 
SMC0190 & SMC\_SC4 & 193064 &    0.4 &    1.28378 &  619.38448 &    17.781 &    17.094 & 0.094 & FU\\ 
SMC0191 & SMC\_SC4 & 192867 &    1.2 &    3.69281 &  616.72477 &    16.061 &    15.418 & 0.094 & FU\\ 
SMC0197 & SMC\_SC5 & 170398 &    1.2 &    0.84209 &  464.19162 &    17.396 &    16.817 & 0.101 & DM\\ 
SMC0198 & SMC\_SC5 & 190473 &    1.2 &    5.74893 &  464.56259 &    16.363 &    15.448 & 0.101 & FU\\ 
SMC0205 & SMC\_SC5 & 288813 &    0.6 &    1.92043 &  464.31693 &    16.995 &    16.305 & 0.101 & FU\\ 
SMC0210 & SMC\_SC6 & 122308 &    0.8 &    3.19828 &  463.27989 &    16.669 &    15.961 & 0.094 & FU\\ 
SMC0216 & SMC\_SC6 & 296732 &    1.3 &    2.16626 &  462.86401 &    16.300 &    15.659 & 0.094 & FO\\ 
SMC0216 & SMC\_SC6 & 296748 &    1.0 &    1.84556 &  463.27691 &    17.558 &    16.737 & 0.094 & FU\\ 
SMC0216 & SMC\_SC6 & 296846 &    1.4 &    1.36357 &  463.75244 &    17.305 &    16.535 & 0.094 & FO\\ 
SMC0225 & SMC\_SC7 & 236366 &    0.6 &    1.14305 &  619.93406 &    18.205 &    17.509 & 0.097 & FU\\ 
SMC0233 & SMC\_SC9 & 163578 &    0.9 &    5.09339 &  615.80084 &    15.890 &    15.179 & 0.076 & FU\\ 
SMC0237 & SMC\_SC10 &  41744 &   1.5 &    1.84743 &  618.57129 &    17.230 &    16.606 & 0.079 & FU\\ 
NGC 346 & SMC\_SC8  & 160784 &   --  &    1.10730 &  619.49347 &    17.354 &    16.770 & 0.100 & FO\\
}

Cluster Cepheids were extracted from the Catalog of Cepheids when the location 
of a given object on the sky was smaller than 1.5 radius from the center of a 
given cluster. Beside classical Cepheids listed in the Catalogs also 
double-mode and second overtone objects from the SMC (Udalski \etal 1999a,b) 
and LMC (Udalski \etal in preparation) were checked. 

204 and 132 Cepheids in the LMC and SMC, respectively, satisfied our 
criterion. Basic parameters of these objects are given in Tables~1 and 2. 
First column is the cluster designation according to the OGLE scheme. 
Cross-identification of clusters with other catalogs can be found in Catalogs 
of Star Clusters. Star ID number (OGLE identification: field and number) and 
distance from the cluster center, measured in units of cluster radius are 
given in columns 2, 3 and 4. Periods, zero phases corresponding to maximum 
brightness, {\it VI} photometry, interstellar reddening and classification 
taken from the Catalogs of Cepheids (Udalski \etal 1999c,d) are presented in 
the following columns. FU, FO, BR and FA symbols in the last column indicate 
that a given object belongs to the classical Cepheids pulsating in the 
fundamental mode, first overtone mode, it is brighter than FO or fainter than 
FU, respectively. DM indicates double mode Cepheid while SO -- second overtone 
object. For the sake of completeness we additionally included to the SMC list 
one Cepheid which is likely a NGC346 member. Large part of that cluster is 
located outside the OGLE-II fields and therefore it was not included 
in the Catalog of Star Clusters from the SMC. 

\vspace*{-5pt}
\Section{Conclusions}
\vspace*{-5pt}
We present lists of Cepheids located in the close neighborhood of star 
clusters from the 4.5 square degrees field of the LMC and 2.4 square degrees 
area of the SMC. Thus far, presented Cepheids constitute the most complete 
sample of such objects with homogeneous observational data and high 
statistical completeness. The sample is very well suited for further detailed 
studies. Results of analysis of these objects will be presented in separate 
papers. 

Photometry of Cepheids and star clusters in the LMC and SMC is available
from the OGLE Internet archive: {\it http://www.astrouw.edu.pl/\~{}ogle} \\
or its US mirror {\it http://www.astro.princeton.edu/\~{}ogle}

\Acknow{The paper was partly supported by the KBN grants: 2P03D00814 to A.\ 
Udalski and 2P03D00617 to G.\ Pietrzy{\'n}ski. Partial support for the OGLE 
project was provided with the NSF grant AST-9820314 to B.~Paczy\'nski.}

\end{document}